\documentclass[twocolumn,prl,preprintnumbers,amsmath,amssymb,longbibliography]{revtex4-2}
\usepackage[pdftex]{graphicx}
\usepackage{graphicx}
\usepackage{epstopdf}
\usepackage{dcolumn}
\usepackage{bm}
\usepackage[utf8]{inputenc}
\usepackage[usenames,dvipsnames]{color}
\usepackage{hyperref}
\usepackage{natbib}
\usepackage[normalem]{ulem}
\usepackage[super]{nth}

\begin{document}

\def\black{\textcolor{black}}
\def\br{\textcolor[rgb]{0.7,0.5,0}}
\def\prp{\textcolor[rgb]{0.5,0,0.5}}
\def\blau{\textcolor{black}}
\def\gr{\textcolor{green}}

\def\StructGrapheneAbs{$7\sqrt{3}\times7\sqrt{3}$}
\def\LDP{LD-phase}
\def\HDP{HD-phase}
\def\AB{{\it A-B}}
\def\Ef{$E_{\rm F}$}
\def\Tc{$T_{\rm C}$}
\def\kpara{{\bf k}$_\parallel$}
\def\kparax{{\bf k}$_{\parallel,x}$}
\def\kparay{{\bf k}$_{\parallel,y}$}
\def\kz{{\bf k}$_\perp$}
\def\kperp{{\bf k}$_\perp$}
\def\Gbar{$\overline{\Gamma}$}
\def\Mbar{$\overline{\rm M}$}
\def\dirGX{$\overline{\rm \Gamma}-\overline{\rm X}$}
\def\dirGY{$\overline{\rm \Gamma}-\overline{\rm Y}$}
\def\dirGK{$\overline{\rm \Gamma}-\overline{\rm K}$}
\def\dirGM{$\overline{\rm \Gamma}-\overline{\rm M}$}
\def\dirGMb{${\rm \Gamma}-{\rm M}$}
\def\dirGRb{${\rm \Gamma}-{\rm R}$}
\def\dirXRb{${X}-{R}$}
\def\dirGMG{$\overline{\rm \Gamma}-\overline{\rm M}-\overline{\rm \Gamma}$}
\def\dirGS{$\overline{\rm \Gamma}-\overline{\rm S}$}
\def\dirXRMXb{${\rm X}-{\rm R}-{\rm M}-{\rm X}$}
\def\dirGN{$\overline{\rm \Gamma}-\overline{\rm N}$}
\def\dirGMtic{$\overline{\rm \Gamma}_{\rm TiC}-\overline{\rm M}_{\rm TiC}$}
\def\dirGKtic{$\overline{\rm \Gamma}_{\rm TiC}-\overline{\rm K}_{\rm TiC}$}
\def\dirMKtic{$\overline{\rm M}_{\rm TiC}-\overline{\rm K}_{\rm TiC}$}
\def\dirGMgr{$\overline{\rm \Gamma}_{\rm Gr}-\overline{\rm M}_{\rm Gr}$}
\def\dirGKgr{$\overline{\rm \Gamma}_{\rm Gr}-\overline{\rm K}_{\rm Gr}$}
\def\dirMKgr{$\overline{\rm M}_{\rm Gr}-\overline{\rm K}_{\rm Gr}$}
\def\dirGMsc{$\overline{\rm \Gamma}_{\rm SC}-\overline{\rm M}_{\rm SC}$}
\def\dirGKsc{$\overline{\rm \Gamma}_{\rm SC}-\overline{\rm K}_{\rm SC}$}
\def\dirMKsc{$\overline{\rm M}_{\rm SC}-\overline{\rm K}_{\rm SC}$}
\def\dirGNhalf{$\frac{1}{2}(\overline{\rm \Gamma}-\overline{\rm N})$}
\def\pntG{$\overline{\rm \Gamma}$}
\def\pntM{$\overline{\rm M}$}
\def\pntK{$\overline{\rm K}$}
\def\pntN{$\overline{\rm N}$}
\def\pntGsc{$\overline{\rm \Gamma}_{\rm SC}$}
\def\pntKsc{$\overline{\rm K}_{SC}$}
\def\pntMsc{$\overline{\rm M}_{SC}$}
\def\pntGtic{$\overline{\rm \Gamma}_{TiC}$}
\def\pntKtic{$\overline{\rm K}_{\rm TiC}$}
\def\pntMtic{$\overline{\rm M}_{\rm TiC}$}
\def\pntGgr{$\overline{\rm \Gamma}_{\rm Gr}$}
\def\pntKgr{$\overline{\rm K}_{\rm Gr}$}
\def\pntMgr{$\overline{\rm M}_{\rm Gr}$}
\def\pntNhalf{ $\overline{\rm N}/2$ }
\def\invA{\AA$^{-1}$}
\def\DCgamma{${\rm DC}_{\overline{\Gamma}}$}
\def\DCNhalf{${\rm DC}_{\overline{\rm N}/{\rm 2}}$}
\def\root33{$\sqrt{3}\times\sqrt{3}$ {\it R}30$^\circ$}
\def\RT3{$\sqrt{3}$}
\def\aR{\alpha_{\rm R}} 

\def\twobytwo{$2\times2$}
\def\twotimestwo{$2\times2$}

\def\CPB{CsPbBr$_3$}
\def\CsPbI3{CsPbI$_3$}
\def\CsPbBr3{CsPbBr$_3$}
\def\CsPbCl3{CsPbCl$_3$}
\def\MAPbI3{MAPbI$_3$}
\def\MAPbBr3{MAPbBr$_3$}
\def\MAPbCl3{MAPbCl$_3$}
\def\MAPbX3{MAPb$X_3$}
\def\CsPbX3{CsPb$X_3$}
\def\mh{$m^*_{\rm h}$}
\def\me{$m_0$}

\title{Is There a Polaron Signature in Angle-Resolved Photoemission of \CPB ? }
 
\author{Maryam Sajedi,$^{1,2}$ Maxim Krivenkov,$^{1}$  Dmitry Marchenko,$^{1}$ Jaime S\'anchez-Barriga,$^1$ Anoop K. Chandran,$^{3,4}$ Andrei Varykhalov,$^1$ Emile D. L. Rienks,$^1$ Irene Aguilera$^{5}$}
\altaffiliation [Present address: ] {Institute of Physics, University of Amsterdam, Science Park 904, 1098 XH Amsterdam, Netherlands.}
\author{Stefan Bl\"ugel$^{3}$} 
\author{Oliver Rader$^1$ }

\affiliation{$^1$ Helmholtz-Zentrum Berlin f\"ur Materialien und Energie, Albert-Einstein-Str. 15, 12489 Berlin, Germany}
\affiliation{$^2$ Institut f\"ur Physik und Astronomie, Universit\"at Potsdam, Karl-Liebknecht-Str. 24/25, 14476 Potsdam, Germany}
\affiliation{$^3$ Peter Gr\"unberg Institut and Institute for Advanced Simulation, Forschungszentrum J\"ulich and JARA, 52425 J\"ulich, Germany}
\affiliation{$^4$ Department of Physics, RWTH Aachen University, 52056 Aachen, Germany}
\affiliation{$^5$ Institute of Energy and Climate Research, IEK-5 Photovoltaics, Forschungszentrum J\"ulich, 52425 J\"ulich, Germany}

\begin{abstract}

The formation of large polarons has been proposed as reason for the high defect tolerance, low mobility, low charge carrier trapping, and low nonradiative recombination rates of lead halide perovskites.
Recently, direct evidence for large-polaron formation has been reported from a 50 \%\ effective mass enhancement in angle-resolved photoemission of \CPB\ over theory for the orthorhombic structure.
We present in-depth band dispersion measurements of \CPB\ and \textit{GW} calculations which lead to almost identical effective masses at the valence band maximum of $0.203 ~\pm 0.016$ \me\ in experiment  
and 0.226 \me\ in orthorhombic theory. We argue that the effective mass can be explained solely on the basis of electron-electron correlation and large-polaron formation cannot be concluded from  photoemission data. 
	
  \end{abstract}

\maketitle

Lead halide perovskites have in recent years become a most important material class for solar cells approaching a tandem-cell efficiency of 30 \%\ \cite{AlAshouri20} as well as light-emitting diodes, semiconductor lasers, and radiation detectors \cite{sum19}.  
They behave very differently from other semiconductors in that they are highly tolerant toward defects \cite{Steirer16,Ball16} while 
showing long carrier lifetimes in the microsecond range and large carrier diffusion length \cite{Biewald19}. The reason for these advantageous properties is, however, not well understood, which makes it difficult to optimize the charge carrier dynamics of this material class further \cite{Herz17,deQuilettes19,Ghosh20}.
The following possibly relevant factors have been reviewed: the influence of trap states on recombination \cite{deQuilettes19,JinH20}, polarons \cite{Herz17,deQuilettes19,Ghosh20}, combinations of the two \cite{deQuilettes19}, ferroelectric domains and boundaries \cite{Ghosh20}, Rashba effect \cite{Ghosh20}, and photon recycling \cite{deQuilettes19}. 

Carrier mobilities turn out to be characteristically low. For single-crystal \CPB, a mobility of 143 cm$^2/$(Vs) has been measured at room temperature \cite{ZhangH17} which, for comparison, is 5 times smaller than for the inorganic semiconductor GaN employed in light-emitting diodes \cite{Goetz98}.
To explain this property, electron-phonon interaction has been analyzed with the result that initially considered acoustic phonon scattering \cite{Shi519}
 is negligible in
methylammonium lead iodide (\MAPbI3) and \CsPbI3 and the  dominant contributions are from longitudinal optical (LO) phonons, which display very similar modes in the two materials \cite{Ponce19}. 
Since these are \blau{ionic solids}, it has been  suggested that large polaron should form at intermediate electron-phonon coupling strength  
and  could explain the combination of four properties \cite{ZhuXY15}: 
the long carrier diffusion length and long lifetimes at modest mobilities, electron-hole recombination rates as low as in 
single-crystal inorganic semiconductors, a temperature dependence of the mobility $\mu\propto T^{-3/2}$, and exceptionally low carrier scattering rate also for hot carriers \cite{Joshi19}.
 The concept of a large-polaron refers to its extent over several lattice sites and is associated with a coherent bandlike transport, the mobility of which decreases	with increasing temperature.
The effect on the mobility is intensively being discussed with a recently predicted reduction by a factor of 2 at low temperature and an enhancement at high temperature \cite{ZhengFan19}.
The idea to connect polaron formation with recombination rates is based on a short-range repulsion of oppositely charged large polarons \cite{Emin18}. 

When electron-phonon coupling is stronger, small polarons should form. These are associated with hopping-type transport and would lead to increasing mobility with increasing temperature.  While the formation of small polarons has been predicted by density functional theory calculations \cite{Neukirch16}, 
$\mu(T)$ data have been interpreted in favor of large-polaron formation \cite{Bonn17,Herz17}. 
An early estimate of the  effective  mass of charge carriers  from the measured mobility and the mean free path  from the momentum relaxation time  arrived at room-temperature values   in the range of 10-300 electron masses \me\ \cite{ZhuXY15}.

The conditions for polaron formation have been investigated in several experiments. Zhu \textit{et al.} \cite{ZhuHScience16,Miyata17}  conducted time-resolved photoluminescence experiments and 
found hot fluorescence emission with $~100$ ps lifetimes in  \MAPbBr3\ but not in \CsPbBr3. It was  
 concluded that the organic cation is crucial and dynamic screening protects energetic carriers  via large-polaron formation. 
 Time-resolved two-photon photoemission and Kerr-effect measurements indicate that large polarons form
due to the deformation of the PbBr$_3^-$ framework, irrespective of the cation type \cite{Miyata17,Miyata17review}. 

Frost \textit{et al}. \cite{Frost17} theoretically investigated the thermalization kinetics for above-band-gap photoexcitation. 
The authors modeled effective masses of (transient) excitons and polarons in \MAPbI3. They calculated for \MAPbI3\  the polaron coupling constants  and
estimated the polaron size based on the Fr\"ohlich model as 26.8 \AA\ (electron) and 25.3 \AA\ (hole) at room temperature. They concluded that the  one order of magnitude higher thermal conductivity of \CsPbI3\  as compared to \MAPbI3\ leads to faster cooling of polarons \cite{Frost17}. 
\black{{\it Ab initio} calculations of transport including multiphonon coupling have recently shown good agreement with transport experiments where also the difference between thick single crystals and thin polycrystalline films of MAPbI$_3$ has been resolved through consideration of photon recycling \cite{Xia21}.}
Using {\it ab initio} many-body \textit{GW} calculations \black{with and without this multiphonon coupling}, Ref. \cite{schlipf18} predicts that polarons in the hybrid perovskite  \MAPbI3\ cause a mass enhancement of 28 \%. Such effective mass renormalization could be tested by angle-resolved photoelectron spectroscopy (ARPES).

   ARPES is a powerful tool to investigate the electronic band dispersion of charge carriers and the interaction with excitations such as phonons \cite{Huefner03}.
For example, in the polar
compound anatase TiO$_2$ where transport is polaronic \cite{Jacimovic12}, photoemission spectral weight is transferred
from the bulk quasiparticle band to photoemission satellites \cite{Moser13}. 
These satellites were described by phonon energies $\hbar \omega_0 = 108$ meV in agreement with a measured LO phonon mode. Second, the effective mass was determined as $0.7$ \me, as compared to the band mass $0.42$ \me\ from density-functional theory (DFT), meaning a mass renormalization of 70 \%\ \cite{Moser13}. 
Thirdly, the polaron contributions to the electron spectral function, including quasiparticle properties and satellites, have been calculated by a cumulant approach \cite{VerdiNC17}. By comparison to calculations without electron-phonon coupling, a theoretical mass enhancement of 73 \%\ was obtained \cite{VerdiNC17}, in good agreement with the ARPES experiment \cite{Moser13} and a large-polaron radius of 57 \AA. 

Also for the surface electronic structure of SrTiO$_3$, which is a polaronic conductor as well \cite{Mechelen08}, 
a distinct satellite feature is seen in ARPES and two satellites can be fitted with  phonon energies of 90 \cite{ChenNC15} to 100 meV \cite{WangNM16}.

 Recently, Puppin \textit{et al.} \cite{Puppin} investigated polaron formation in 
 \CsPbBr3\ by ARPES. The valence band maximum (VBM) at the \textit{R} point was determined by tracing binding energy and effective mass with photon energy, i. e., with the electron wave vector perpendicular to the surface, \kperp. Comparison with DFT for orthorhombic \CsPbBr3\ of 0.17 \me\ (0.12 \me\ for cubic) showed that the measured effective hole mass of 
 $m^*_{\rm h} = 0.26$ \me\ is enhanced by 50 \%\, which was interpreted as direct experimental evidence that charge carriers form large polarons \cite{Puppin}. 

For the present Letter, we have investigated \CsPbBr3\ by ARPES experiments and by \textit{GW} calculations. We obtain experimental and theoretical results very different from the previous work and we discuss possible reasons. In fact, experimental and theoretical effective masses are very similar and do not indicate large-polaron formation. 

\CsPbBr3\ single crystals have been grown by the antisolvent vapor-assisted crystallization  method, adapted from Ref. \cite{RakitaCG16}. 
 \black{ARPES} has been performed with a hemispherical Scienta R8000 analyzer at the ARPES-1$^2$ and RGBL-2 instruments at BESSY-II using linearly polarized undulator radiation. Experiments have been done with freshly cleaved surfaces at room temperature as described previously \black{(see Ref.} \cite{SajediPRB20} \black{and its Supplemental Material for details on sample growth and the ARPES experiment)}. The base pressure of the instruments was better than $2\times10^{-10}$ mbar \black{and the spatial resolution in ARPES $\sim100$ $\mu$m}.
For our calculations, we used the one-shot \textit{GW} method ($G_0 W_0$) as implemented in the all-electron {\sc fleur}{}~\cite{fleur} and {\sc spex}{}~\cite{friedrich10} codes, fully taking into account the spin-orbit coupling~\cite{aguilera13} \black{(see Supplemental Material} \cite{Supplement}, \black{which contains Refs. \cite{Perdew96,koelling77,Li90,mostofi08,Lejaeghere16,osti_calc}).} 

\begin{figure}[t]
\centering
\includegraphics[width=0.48\textwidth]{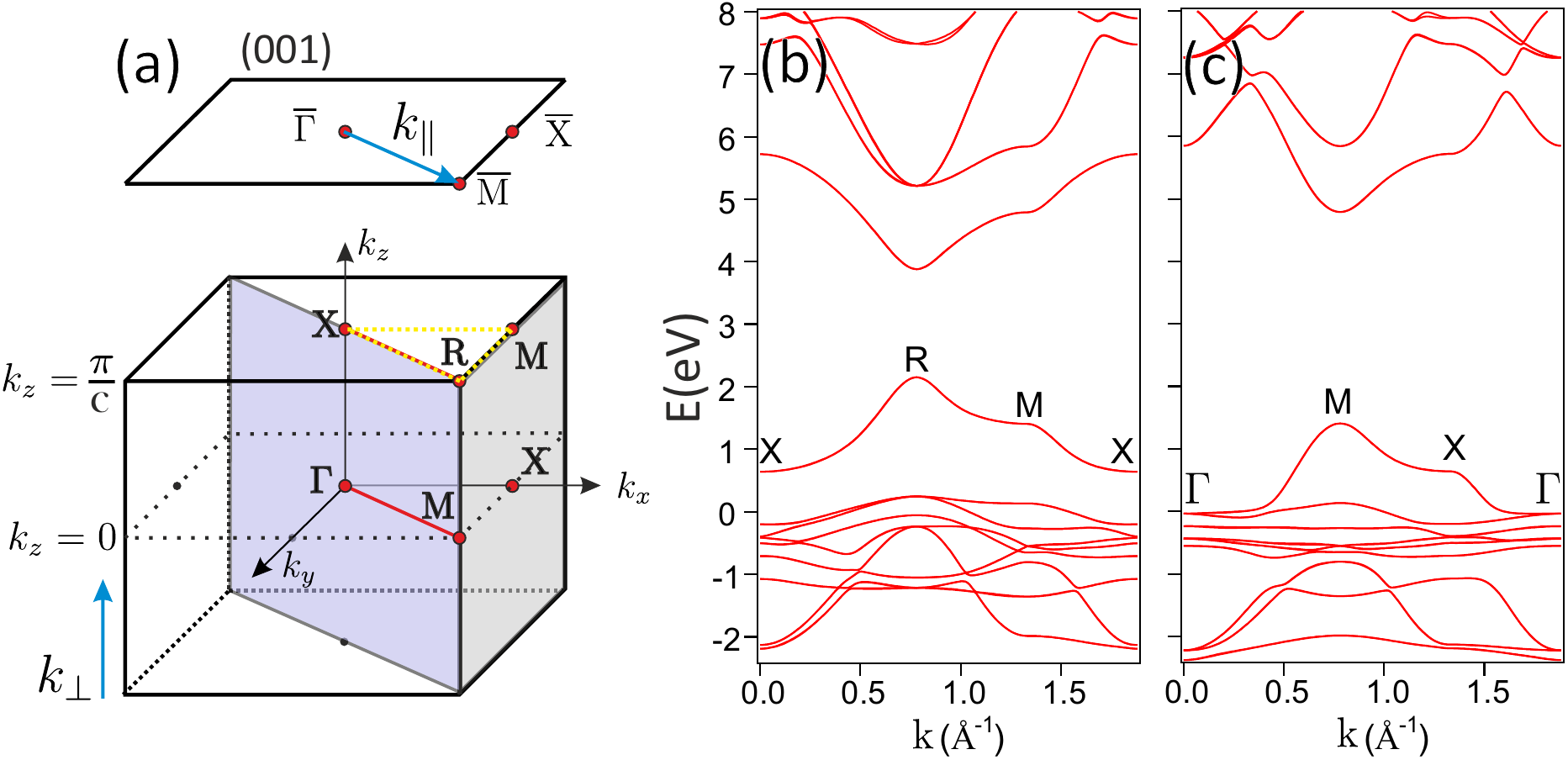}
\caption{\label{Fig.1.} (a) Cubic bulk and surface Brillouin zones, bulk paths \dirGMb\ and \dirXRb\ projected onto \dirGM\ in the surface Brillouin zone. Yellow dashed line shows the \dirXRMXb\ path. (b),(c) Calculated \textit{GW} band dispersions of \CsPbBr3 for the cubic phase presented for {\bf k}$_z$ = $\pi$/c and {\bf k}$_z$ = 0 planes, respectively. See Fig. S4 in the Supplemental Material for the orthorhombic bands.}
\end{figure}

Figure 1 shows the calculated \textit{GW} bands for the cubic structure. \black{Structural details such as lattice parameter of 6.017 \AA\ are taken from \cite{osti_cub}}. Although \CsPbBr3\ is orthorhombic at room temperature (tetragonal above 361 K and cubic above 403 K \cite{rodova03,hirotsu74,ZhangRSC17}), ARPES measurements at room temperature follow the band dispersion for the cubic phase \cite{Puppin}.
The VBM  is situated at the \textit{R} point which is formed by antibonding orbitals, typical of the lead halide perovskites \cite{Frost14,Goesten18}. 
\CsPbBr3\ has a large experimental band gap of around 2.37 eV at room temperature \cite{Mannino20}.
Band gaps are not appropriately described by  DFT  in the local-density or generalized-gradient approximations. Even with the more advanced HSE06 hybrid functional, a band gap of only about 1.17 eV is obtained for \CsPbBr3\ \cite{Kang18}.
We use, therefore, many-body perturbation theory in the \textit{GW} approximation to calculate quasiparticle self-energy corrections for
the electronic states, which yields results that are directly comparable with ARPES measurements. These do not include electron-phonon coupling and
are therefore well suited as bare bands in the search for polaronic mass enhancement. 
The band gap we obtain for the cubic structure is 1.74 eV (Fig. 1). 
Note that \textit{GW} calculations in two previous works \cite{Puppin,wiktor17} lead to smaller band gaps of 1.05 and 0.94 eV, respectively. We tackle the question for this difference in our benchmarking calculations in the Supplemental Material \cite{Supplement}, which contains Refs. \cite{osti_cub, rajeswarapalanichamy20, ye15,ghaithan20, qian16,sun20}.

\begin{figure}[!t]
	\centering
\includegraphics[width=0.45\textwidth]{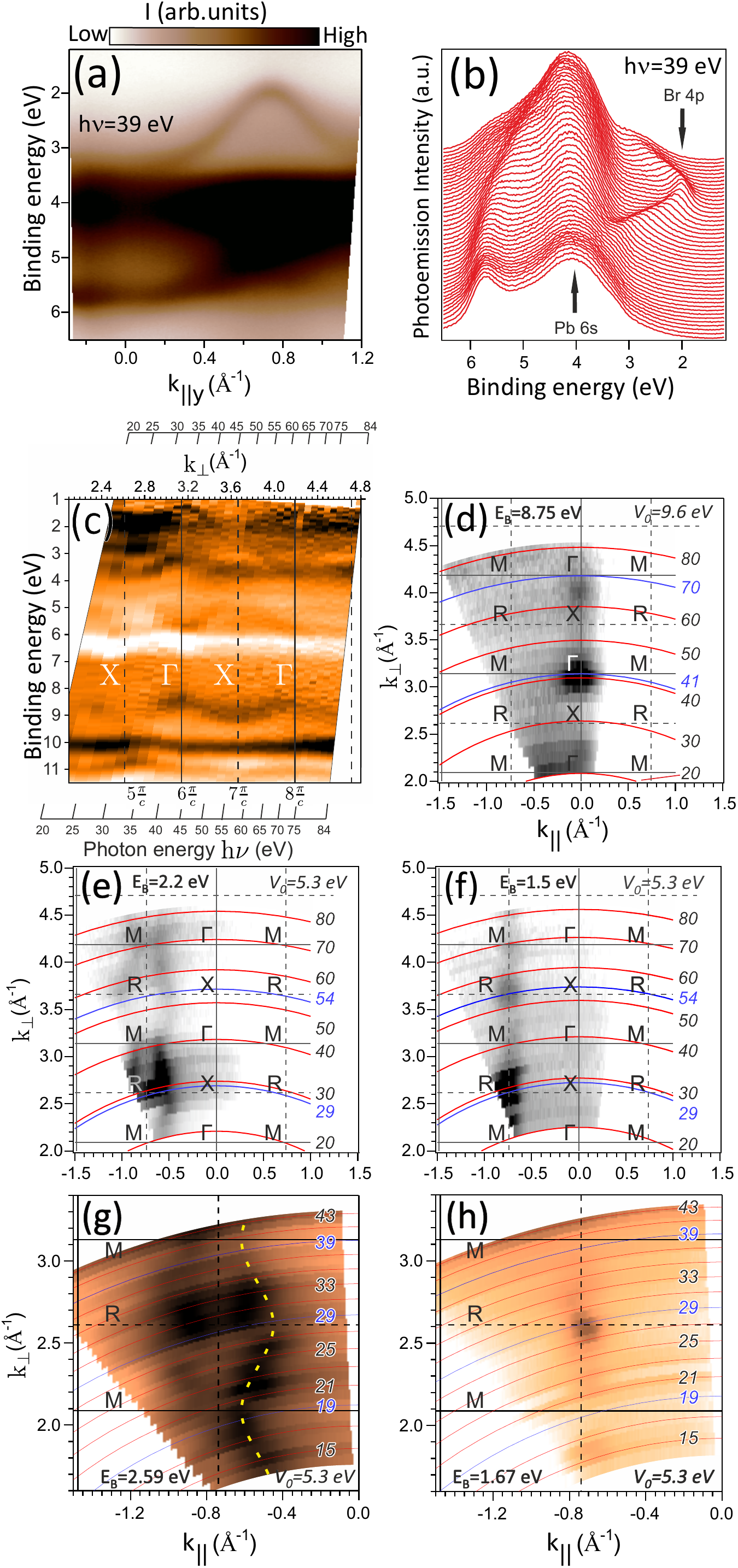}
\caption{\label{Fig.2.} ARPES determination of the photon energy that corresponds to the valence band maximum. (a,b) ARPES at $h\nu = 39$ eV. (c) Photon energy-dependent measurements taken at normal emission ($ {dI}/{dE}$), converted to momentum space corresponding to the $\Gamma-{X}-\Gamma$ bulk direction. (d)--(h) ARPES intensity plots in momentum ($k_{\perp}$, $k_{\parallel}$) space extracted from photon-energy-dependent measurements (in eV) along \dirGM\ 
	for (d) $V_0 = 9.6$ eV, indicating dispersions around $\Gamma$ point, and (e),(f) $V_0 = 5.3$ eV near \textit{R} point for two different binding energies (${E_B}$). (g),(h) Intensity plots in ($k_{\perp}$, $k_{\parallel}$) extracted from another $h\nu$-dependent experiment for a different sample for two binding energies around the first \textit{R} point. (a)--(f) Measured with horizontal and (g),(h) circular light polarization.}
\end{figure} 

Figures 2(a) and 2(b) show overview ARPES spectra of \CsPbBr3 (100). Pb $6s$ states with little dispersion, as well as Br $4p$ states forming the VBM are marked by arrows.
Figure 2(c) displays normal-emission measurements (i. e., at \Gbar) as a function of photon energy between 
	 $h\nu = 21$ and 83 eV, with 2 eV steps. 

These data show a symmetric behavior about the wave vector \kperp = 3.66~\AA$^{-1}$.   
Since the VBM is not expected to occur in normal emission (\Gbar), we can only determine critical points $\Gamma$ and  from extrema in the band dispersion. 
We distinguish \textit{X} from $\Gamma$ through the dispersion of the Pb $6s$-Br $4p$ states \cite{Goesten18} at around 9 eV binding energy, which are running downward from $\Gamma$ to \textit{X}. Our assignment of \kperp = 3.66~\AA$^{-1}$  to the \textit{X} point in the simple model of free-electron final states corresponds to an inner potential of $V_0 = 9.6$ eV. 

The location of the \textit{R} point (projected on \Mbar) is directly revealed by a {\bf k}$_\parallel$-dependent plot at 1.5 eV binding energy, which is near the energy of the VBM [Fig. 2(f)].

 We also show a second binding energy of 2.2 eV [Fig. 2(e)] to confirm that the contours around the VBM close in at 29 and 54 eV photon energies, which are therefore assigned as the \textit{R} point [see the second derivative plot of Fig. 2(e) in the Supplemental Material \cite{Supplement}].
 
This corresponds to a slightly different inner potential of 5.3 eV. \textcolor{black}{In order to verify the reproducibility of VBM selection, we have repeated the photon-energy-dependent experiments for the range of 14-43 eV with 1 eV steps. We demonstrate the {\bf k}$_\perp$ dispersions around the first \textit{R} point ($h\nu = 29$ eV) at two binding energies (2.59 and 1.67 eV) in Figs. 2(g) and 2(h).} 

 \begin{figure*}[!ht]  
	\centering
	\includegraphics[width=0.85\textwidth]{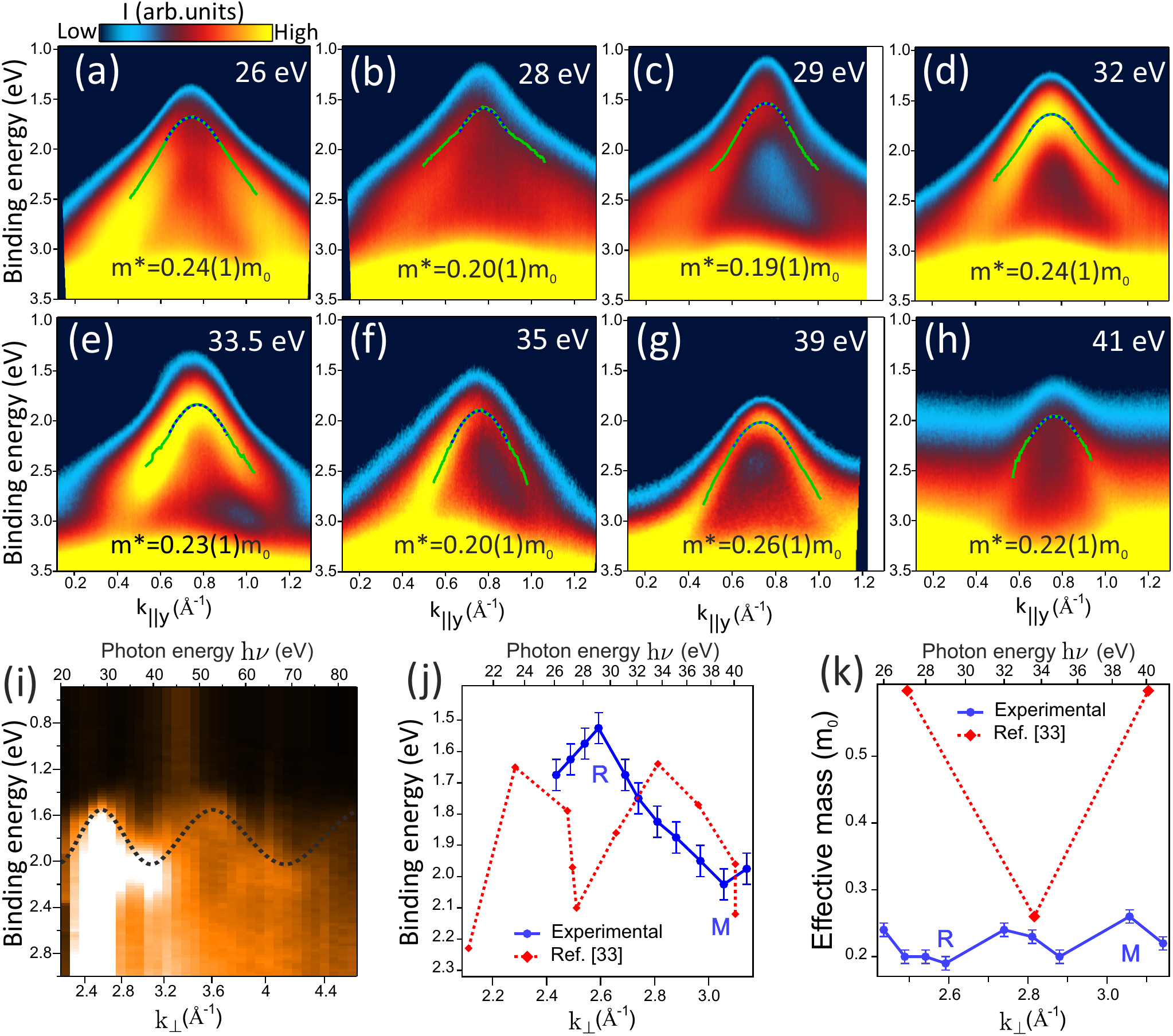}
	\caption{\label{Fig.3.}(a)--(h) ARPES data along the \dirGMG\ direction for various photon energies. Numbers at the right indicate the corresponding photon energies. (i) Photon-energy dependence taken at \Mbar\ point (dashed line guides the eye). (j) Energy of local maxima and (k) effective mass  \mh\ vs. photon energy and {\bf k}$_\perp$ compared to Ref. \cite{Puppin}. {\bf k}$_\perp$ axis is converted from photon energies using $V_0 = 5.3$ eV. \textcolor{black} {The binding energy of the topmost band in (j) has been determined from the individual spectra to have better precision than taking it from the overall {\bf k}$_\perp$ dependence (i).}}
\end{figure*}	
	
In Fig. 3 we show the band dispersion along \dirGM\ for various photon energies.
At first, the binding energy of the local maximum (which is related to the plots in Fig. 2) is evaluated and confirms $h\nu = 29$ eV as the VBM.
Moreover, it is seen that the dispersion at 29~eV appears narrower than those at the other photon energies. This is confirmed by the fits of dispersions and effective masses [Fig. 3(a)--3(h)]. We have repeated the measurement at 29~eV for three more samples \black{(see Fig. S1, Supplemental Material \cite{Supplement})} and obtained \mh\ values of altogether 0.19, 0.23, 0.20, and 0.19 $m_0$, giving an average of \black{$0.203 \pm 0.016$ \me\ (standard deviation)}.   
It should be noted that the previous value of 0.26 \me\ at $h\nu = 33.5$ eV \cite{Puppin} connects to our measurement [0.23 \me\, see Fig. 3(k)],
but the assignment to \textit{R} does not. Our different values for $V_0$ for final states at \Gbar\ and \Mbar\ shift the expected \textit{R} point by only $\sim 2.5$ eV in photon energy and do not explain the different assignment. It appears that the previous assignment of $m^*_{\rm h} = 0.26$ \me\  is rather a consequence of higher \mh\ values  $>0.5$ $m_0$ at 27 eV and 40 eV (Fig. S5 of Ref. \cite{Puppin}), which we cannot confirm (we obtain 0.24 $m_0$ at 26 eV, 0.26 $m_0$ at 39 eV, 0.22 $m_0$ at 41 eV).  
\textcolor{black} {The identification of high-symmetry points corresponds also well with the overall $\rm k_{\perp}$ dispersion taken at \Mbar, see Figs. 3(i) and 3(j).}

We compare the measured effective mass with our \textit{GW} calculations. From our calculations, we obtain $m^*_{\rm h} = 0.15$ \me\ (gap 1.74 eV) for the cubic structure at \textit{R} evaluated along \dirXRb, which is smaller than our ARPES value. 
It has been shown that the orthorhombic structure leads to a larger effective mass than the cubic one \cite{Puppin}. By DFT using the HSE functional, 0.12 \me\ was obtained for cubic and 0.17 \me\ for the orthorhombic structure used to derive the mass enhancement of 50 \%\ \cite{Puppin}.
 Even though the band gap was underestimated by the calculation (1.17 eV), its effect on the effective mass was found to be less than
10 \%\ \cite{Puppin}. A similar behavior was reported by Becker \textit{et al.} \cite{Becker18}. 
Comparing DFT results for different anions \cite{Kang18} or \textit{GW} calculations for the same anion at different
sophistication levels \cite{Ponce19} shows instead that larger gaps have larger effective masses
(see \black{Fig. S3}, Supplemental Material \cite{Supplement}, \black{which also includes Refs. \cite{Kang18,Puppin,filip15}}). 
To check for a similar trend, we also conducted \black{a calculation for the orthorhombic unit cell} and obtained \black{a hole effective mass value of $m^*_{\rm h} = 0.226$ \me\, averaged over the different crystallographic directions
at a band gap  of 2.09 eV, which is closer to the experimental value of 2.37 eV \cite{Mannino20} than other calculations (see Figs. S4, and S5 in the Supplemental Material \cite{Supplement}}). 
This \mh\ value is even slightly larger than our ARPES result. \blau{Because Wannierization was only performed for the valence band, there is the possibility that the effective mass is slightly overestimated in the orthorhombic calculation. 
We believe that such overestimation is, on the other hand, rather limited. Note that, in our cubic calculation, as well as in that of Ref. \cite{Puppin}, Wannierization was performed for valence and conduction bands and still the effective masses differ
substantially (0.15 $m_0$ in the present case and 0.12 $m_0$ in Ref. \cite{Puppin}). 
We conclude that} the measured effective mass can be explained by electron-electron correlations 
included in our theoretical description and does not indicate large-polaron formation. 

As pointed out before \cite{Puppin}, the LO phonon energy of \CsPbBr3\ is very small ($\le25$ meV) and satellites in 
ARPES should be difficult to observe.
This raises the question whether lead halide perovskites are the right  model systems for the observation of polaron formation and tin halide perovskites with phonon energies of more than 180 meV in CsSnBr$_3$ \cite{HuangLy13} are not suited better. \textcolor{black}{DFT calculations \cite{meggiolaro20} indicate a rise in the polaron stabilization energies upon exchange of Pb by Sn in \MAPbI3.} 

We showed that ARPES provides high-quality band structure data of halide perosvkites and helps assessing relevant physics for lead halide perovskites.  Recently, we demonstrated by ARPES that a large valence band Rashba effect is absent in MAPbBr$_3$ and \CsPbBr3\ \cite{SajediPRB20}. We further note that small-polarons cannot be detected through the effective mass.   
While  small polarons can cause deterioration of device performance by charge trapping \cite{Neukirch16,meggiolaro20}, 
it has recently been pointed out that in \CsPbBr3\ small polarons can also reduce carrier recombination \cite{Osterbacka20}.
The experimental temperature-dependent charge mobility of lead halide perovskites 
 was found not to be as supportive for the Fr\"ohlich model of large polarons  as was previously believed, and also compatible with small polaron formation \cite{Osterbacka20}.
Moreover, recent calculations suggest that polaronic effects on scattering and mobilities of charge carriers are more limited than previously claimed \cite{Irvine21}.

 In conclusion, our experimental hole effective mass lies in between the results of our \textit{GW} calculations, and 
\black{despite recent success of Fr\"ohlich polaron-based models for transport data}, we do not find any significant mass enhancement, regardless of whether we compare to the cubic or the orthorhombic calculation. 
Our results are based on two insights: First, a precise
determination of the point in momentum space is necessary for extracting the correct effective mass
from the experiment. Second, the determination of the theoretical effective mass requires a calculation
that does not strongly underestimate the band gap. 
The supporting data for the calculations in this Letter are openly available from the Materials Cloud repository \cite{Sajedi22database}.

\section{Acknowledgements}
We thank Professor Thomas Kirchartz for helpful discussions and continuous support and Professor Eva Unger for drawing our attention to 
\CsPbBr3.
Use of the  Helmholtz Innovation Lab HySPRINT for sample preparation is gratefully acknowledged as well as 
 computing time granted through JARA-HPC on the supercomputer JURECA at Forschungszentrum J\"ulich.
 J. S.-B. gratefully acknowledges financial support from the Impuls- und Vernetzungsfonds
 der Helmholtz-Gemeinschaft under Grant No. HRSF-0067.

\phantom{xxxx}

\bibliography{library_polaron} 

\end{document}